\documentclass[sigconf]{acmart}

\setcopyright{none}
\acmConference[WWW '27]{the ACM Web Conference 2027}{May 10--14, 2027}{Dublin, Ireland}
\acmYear{2027}
\copyrightyear{2027}
\acmDOI{}
\acmISBN{}
\renewcommand\footnotetextcopyrightpermission[1]{}
\usepackage{booktabs}
\usepackage{graphicx}
\usepackage{float}
\usepackage{cuted}
\usepackage{multirow}
\usepackage{xspace}
\usepackage{array}
\usepackage{pifont}
\newcommand{\system}{EvoCache\xspace}

\newcommand{\rwyes}{\ding{51}}
\newcommand{\rwout}{--}

\title{EvoCache: Accelerating Browser Large Language Model Delivery with Public Cache Anchors}

\author{Zhiyu Wang}
\email{zhiyu.wang4@unimelb.edu.au}
\affiliation{%
  \institution{The University of Melbourne}
  \city{Melbourne}
  \country{Australia}}

\author{Rajkumar Buyya}
\email{rbuyya@unimelb.edu.au}
\affiliation{%
  \institution{The University of Melbourne}
  \city{Melbourne}
  \country{Australia}}

\begin{document}

\begin{abstract}
Large Language Models (LLMs) can run directly in a browser, but first use typically requires downloading model files that are hundreds of MiB or even several GiB in size. The same LLM is often repackaged by different publishers; these files share substantial content, yet byte differences prevent whole-file caches from reusing them. We propose \system, a browser-managed LLM delivery system. \system uses public cache anchors that are stored in advance by the browser and independent of browsing history to reconstruct losslessly the publisher version requested by a site. A site specifies only the target, while the browser selects the anchor and performs reconstruction, keeping both hidden from the site. An audit of 1,000 public repositories finds that 61 of the 100 most popular LLMs are offered by multiple publishers and that 98.1\% of overlapping versions cannot be reused by whole-file caching. Across 23 cross-publisher model-package pairs, \system achieves a median traffic reduction of 77.3\% and outperforms every comparator. Across 240 real end-to-end browser experiments, \system reduces network traffic by 78.9\% and page-ready time by 64.2\% relative to ordinary full downloads. For cached data left by other sites, \system also protects users' cross-site cache privacy by using the same return time for cache hits and misses.
\end{abstract}

\ccsdesc[500]{Information systems~Browsers}
\ccsdesc[500]{Computer systems organization~Client-server architectures}
\ccsdesc[300]{Security and privacy~Web application security}
\keywords{in-browser large language models, Web caching, delta compression, cache privacy}

\maketitle

\section{Introduction}

Browser-based Large Language Models (LLMs) allow Web pages to run models directly on users' devices, reducing server costs while keeping sensitive data local~\cite{weinfer-www,nnjit,browser-anatomy}. The browser must first download the model weights~\cite{wamo-www}. Even after quantization, a model package is typically hundreds of MiB, and larger models can reach several GiB~\cite{latune-www}. Model downloads therefore often determine how long users must wait.

Caching should reduce repeated downloads, but different publishers often package the same LLM differently. Even when the model architecture and quantization method are identical, differences in calibration data, tensor types, metadata, and file layout change the final GGML Universal File (GGUF). An audit of 1,000 public model repositories finds that 61 of the 100 most popular LLMs are offered by multiple publishers. Among 738 groups containing the same LLM and quantization configuration from different publishers, 724 (98.1\%) contain no byte-identical files. Ordinary whole-file caching can therefore reuse almost none of this previously downloaded data.

A browser also cannot directly expose one publisher's cached files to other sites. If model installation speed depends on whether a cache entry exists, a site can use timing to infer whether the user installed an LLM on another site. Modern browsers prevent such cross-site observation by partitioning storage~\cite{timing-web,cache-partition}. Existing sharing mechanisms handle only more limited cases. Cross-Origin Storage (COS) requires byte-identical objects~\cite{cos}; Compression Dictionary Transport (CDT) requires a site to specify a dictionary in its partition~\cite{cdt-rfc,chrome-cdt}; and cross-origin model caching likewise depends on identical complete model files~\cite{hybrid-cache}. None allows a site to specify only its desired target while the browser privately reuses another publisher's model package. This gap lies in the Web's own infrastructure: how browsers deliver, cache, and isolate large resources across sites, the focus of the Web Infrastructure and Agentic Systems track.

This paper proposes \system, which places a trusted broker inside the browser and uses public anchors that are provisioned by the browser and independent of site-visit history. A site submits only a publisher-authorized target. The broker matches corresponding tensors and, for each data chunk, selects exact reuse, compressed fetch, or lossless residual reconstruction. The final file must exactly match the target supplied by the publisher. The selected anchor, reconstruction plan, transferred bytes, and execution progress remain hidden from the site. A target reconstructed from a public anchor can be returned immediately. For cached data left by site activity, \system returns results at a predetermined public time, thereby protecting users' cross-site privacy.

We implement \system in a privileged Chromium extension. Across 23 pairs of LLM packages from different publishers that are difficult to reuse directly, \system achieves a median traffic reduction of 77.3\%, 27.9 percentage points more than the best generic file-differencing method given the same tensor-structure information, and leads on every pair. Its minimum traffic reduction remains 76.6\% for larger LLMs. A further 240 real end-to-end browser experiments show that, relative to ordinary full downloads, \system reduces network traffic by 78.9\% and page-ready time by 64.2\%. Relative to the best-performing comparator, it further reduces network traffic by 14.2\% and page-ready time by 7.9\%.

This paper makes three contributions:

\begin{itemize}
  \item Through a large-scale repository audit and byte-level analysis, we show that cross-publisher repackaging is a widespread problem for browser caching and reveal that different GGUF files still contain substantial compressible redundancy at the tensor level.
  \item We design \system. The system selects reusable anchors inside the browser and chooses lossless reconstruction operations according to the characteristics of each data chunk. It both reproduces the publisher-authorized target exactly and keeps cache state and reconstruction hidden from the site.
  \item We implement and evaluate a complete browser delivery path. The experiments cover 45 real model packages, publishers and LLMs not used during system development, larger LLMs, and 240 real end-to-end browser runs, in which \system outperforms every comparator in both page-ready time and network traffic.
\end{itemize}

The rest of the paper is organized as follows. Section~\ref{sec:problem} establishes the caching problem created by cross-publisher repackaging. Section~\ref{sec:privacy} defines the system goals and security rules. Section~\ref{sec:design} presents the design of \system. Section~\ref{sec:eval} evaluates the system. Section~\ref{sec:related} discusses related work. Section~\ref{sec:conclusion} concludes the paper.

\section{Why Whole-File Caching Fails for LLMs}\label{sec:problem}

\subsection{The Tension in Web Caching}

Unpartitioned caching accelerates identical third-party resources but turns cache hits into cross-site history signals, so modern browsers partition network and storage state~\cite{timing-web,cache-partition}. COS shares public-list-admitted, content-addressed files~\cite{cos}. CDT reconstructs a response against a matching cached dictionary under HTTP partitioning and negotiation~\cite{cdt-rfc}; the server learns the selected dictionary, and Chromium caps a dictionary at 100 MiB~\cite{chromium-cdt}. Only 3 of the 45 model packages in the corpus fall below this limit. These mechanisms reuse a complete object or an explicitly negotiated dictionary.

LLM packaging introduces another dimension: the same LLM may be quantized and converted independently by multiple publishers. GGUF stores typed metadata, tensor descriptions, alignment information, and tensor bytes~\cite{gguf}. Metadata order changes file offsets, while importance matrices and per-tensor type selection change the quantized bytes. A target site knows which publisher's model package it needs, but it should not learn which other eligible model packages the browser holds.

\subsection{How Common Is Independent Repackaging?}

We audit the 1,000 most-downloaded GGUF repositories returned by the public Hugging Face API on August 22, 2026, and group them by their declared upstream LLM~\cite{hf-api}. Among the 100 most-downloaded LLMs, 61 are offered by at least two independent publishers, with a median of three publishers. These multi-publisher LLMs account for 74.8\% of downloads among the relevant repositories. Among 738 cross-publisher version groups for the same LLM and quantization configuration, 724 (98.1\%) share no byte-identical Git Large File Storage (LFS) object. This result shows that independent repackaging of the same LLM is common, yet these model packages can almost never directly hit a whole-file cache.

\subsection{Audited Cross-Publisher Corpus}\label{sec:corpus}

Table~\ref{tab:corpus} summarizes 45 GGUF model packages selected from public Hugging Face repositories. Each pair contains versions from two publishers for the same upstream LLM and quantization class, producing 42 directed pairs. The source and target files in every pair are different, so ordinary whole-file caching cannot produce a hit. Corresponding tensors are matched by name, shape, and quantization class to analyze calibration and per-tensor type differences introduced by different publishers.

\begin{table}[t]
  \caption{Real model-package corpus.}
  \label{tab:corpus}
  \centering
  \begin{tabular}{@{}>{\raggedright\arraybackslash}p{0.36\columnwidth}>{\raggedright\arraybackslash}p{0.59\columnwidth}@{}}
    \toprule
    \textbf{Property} & \textbf{Value} \\
    \midrule
    LLM families & SmolLM2-135M; SmolLM2-360M; Qwen2.5-0.5B; Llama-3.2-1B \\
    Publishers / immutable repositories & 4 / 12 \\
    Quantizations & Q3\_K\_M, Q4\_K\_M, Q5\_K\_M, Q8\_0 \\
    Model packages / evaluated pairs & 45 / 42 \\
    Corpus size & 18.4 GiB \\
    Target sizes & 89.2 MiB--1.23 GiB \\
    \bottomrule
  \end{tabular}
\end{table}

After dividing the model packages into chunks, the 42 pairs fall into two groups. In 19 pairs, at least 90\% of chunks are identical and can be reused directly by \system. In the other 23 pairs, hereafter called \emph{calibration-divergent pairs}, different publishers used different calibration settings and identical chunks account for no more than 41.8\%. Most content in these pairs must therefore be reconstructed from residuals, and they are used to evaluate \system's core reconstruction method.

\subsection{Why Quantized Bytes Differ}

Why can two files differ while remaining reusable? Their publishers start from the same logical tensors, but calibration data can change small values inside the packed representation, and importance-aware policies can assign different types to some tensors. Across all 23 calibration-divergent pairs, publishers either differ in whether they use an importance matrix or use different calibration sources and chunk counts. Tensor names and shapes remain the same in every pair. Eleven pairs also retain every tensor type and extent; the other twelve change some type assignments. Within tensor bytes that remain positionally comparable (91.1\% median coverage), 81.0\% are equal at the median and only 5.79\% of bits flip. These sparse, aligned changes explain why compressed residuals remain small even when nearly every 1-MiB chunk differs. \system uses these shared bytes to reconstruct the publisher-selected package while preserving its quantization and calibration choices.

\section{Preventing Cross-Site Cache Probing}\label{sec:privacy}

\subsection{Timing Differences Expose Cache State}

When the browser already holds reusable data, installing an LLM is usually faster than downloading the complete file again. A malicious site can repeatedly request the same LLM and measure the completion time of each request. If cache hits are consistently faster, the site can determine whether the browser holds related data and infer whether the user previously installed the same LLM on another site.

The threat model considers an attacker that controls a Web origin. It can request any authorized target, use high-resolution clocks, and observe API results, errors, and Performance entries available to the page. Normal browser isolation prevents it from directly reading process memory, another origin's OPFS, or browser-internal requests, and the attacker cannot forge the publisher's Ed25519 signature~\cite{ed25519}. It can therefore probe the cross-site cache only through installation behavior visible to the page. We assume publisher endpoints do not collude with the calling origin, and strict release reuses history-derived data only for publishers on a browser-maintained list. \system targets observations exposed through the page's APIs and complements browser-level isolation of shared hardware resources.

\subsection{The Page Can Specify Only the Target}\label{sec:api}

To prevent a page from actively probing the browser cache, \system allows it to specify only an authorized target, not the base package or reconstruction method. The page already knows the target it requested and the model packages it supplied. What must remain hidden is whether the browser found other reusable data, which base package it selected, and the reconstruction plan, transfer volume, and execution progress. The interface is:

\begin{verbatim}
navigator.modelPackages.install({
  root, manifestURL, publisherKey
}) -> ModelPackageHandle
\end{verbatim}

Here \texttt{root} identifies the target, and \texttt{manifestURL} and \texttt{publisherKey} establish publisher authorization; the browser, not the page, selects the release policy described in Section~\ref{sec:release}. The returned \texttt{ModelPackageHandle} opens only the requested target.

The interface exposes no \texttt{has} or \texttt{list} operation, so a page cannot query which LLM files the browser has cached. It also exposes no base-package parameter, byte counter, selected reconstruction method, or fine-grained progress. Base-package selection, planning, download, and reconstruction all remain inside the browser broker.

The broker must also decide whether to accept the request before reading shared cache state. This decision uses only the security policy, publisher authorization, caller-owned quota, and public target-size, device-capacity, and network classes. Shared storage has a separate accounting domain, so its occupancy cannot affect admission. If the browser cannot reserve enough time and resources for a full cold download and commit, it rejects the request before cache lookup. Acceptance or rejection therefore cannot reveal whether reusable data exists.

\subsection{Return Rules for Two Cache Classes}\label{sec:release}

\system defines one return rule for each of the two kinds of reusable data. Public anchors provisioned by the browser contain no browsing history, so their results can be returned as soon as reconstruction finishes. Cache left after another site installs an LLM may reflect browsing history, so the corresponding result can be returned only at a public deadline. \system calls the latter rule \emph{strict release}. The broker chooses the rule from public information before cache lookup: a target with a browser-provisioned anchor uses immediate release and reuses only that anchor, while every other target from a listed publisher uses strict release whether or not history-derived data exists. Targets from unlisted publishers without an anchor reuse only data in the requesting site's own partition.

For a request, the public description $p$ contains the target identity and size together with a public CPU/network service class. Before cache lookup, the broker computes a release time $D(p)$ from the time reserved for a cold download, commit, and a public safety margin. Because this calculation occurs before cache lookup, $D(p)$ does not depend on whether reusable data exists.

A cache hit may finish reconstruction early, but the broker waits until $D(p)$ to return the handle. A cold download returns its handle at the same time. The admission check in Section~\ref{sec:api} ensures that the cold path can finish before $D(p)$. The page therefore observes the same completion time in both cases, while a cache hit still saves network bytes.

Formally, for an admitted request class $p$, let $C_0$ and $C_1$ differ only in cross-origin eligible cache state. When both paths complete by $D(p)$, the page observes
\[
  \begin{aligned}
  T_{\mathrm{API}}(p,C_0)&\equiv T_{\mathrm{API}}(p,C_1)\\
  &=(\mathrm{accepted},\ D(p),\ \mathrm{target\ handle}).
  \end{aligned}
\]
This expression says that both requests are accepted, both return at $D(p)$, and both produce a handle for the requested target. The relation $\equiv$ also requires opaque handle identifiers to have the same distribution in both cases. The broker emits no intermediate API events, and corresponding failures expose the same errors. Appendix~\ref{app:proof} gives the isolation and scheduling conditions needed for these properties. When those conditions hold, the page cannot distinguish $C_0$ from $C_1$ through any API-visible observation. Because $D(p)$ adds a public safety margin to the cold-path estimate for the declared network class, the cold path completes before $D(p)$ whenever the network meets that class; the guarantee therefore covers every request the broker admits under its declared class.

A site-facing API cannot mix the two cache-entry classes or turn a history-dependent entry into a public anchor. Only browser policy can decide that an entry belongs in the public-anchor namespace.

\section{EvoCache's LLM Reconstruction Workflow}\label{sec:design}

\begin{figure*}[t]
  \centering
  \includegraphics[width=\textwidth]{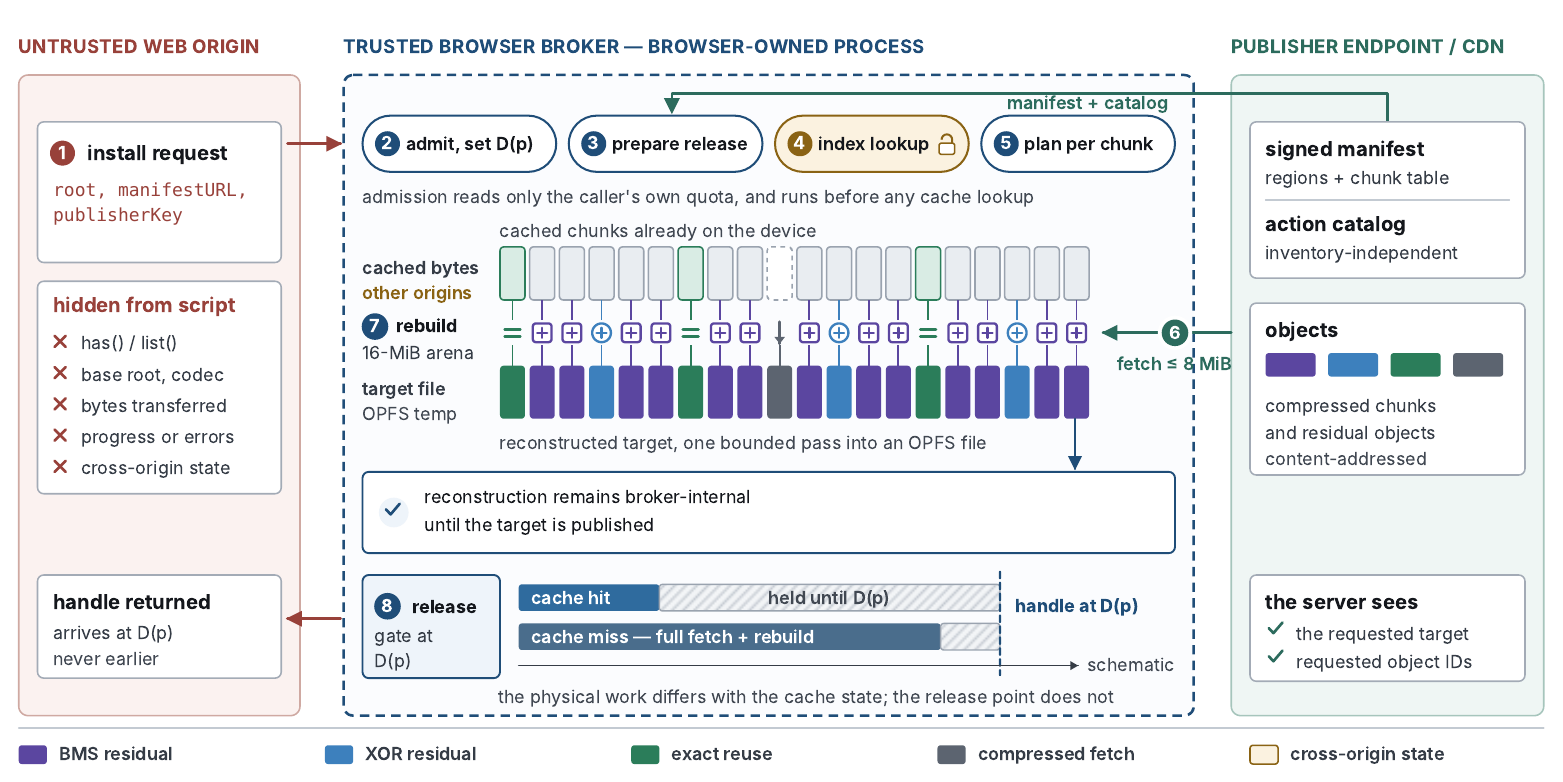}
  \caption{EvoCache's broker boundary and strict-release data path. The calling origin supplies public request metadata and receives a final handle only at the public deadline \(D(p)\). Eligible cached roots, the selected plan, byte counts, and cache-dependent progress remain inside the trusted broker and hidden from the calling origin.}
  \Description{A numbered workflow separates the untrusted Web origin, trusted browser broker, and publisher origin. The broker admits the request, checks publisher authorization, performs a private index lookup, plans one action per chunk, fetches content-addressed objects, rebuilds the target, and releases its handle at a public deadline.}
  \label{fig:architecture}
\end{figure*}

This section explains how \system reconstructs and delivers a target LLM. The broker first verifies the target and aligns reusable model packages, then selects a reconstruction method for each data chunk, and finally reconstructs and verifies the target as a stream before delivering it. \system can reuse two classes of data. The first is a public anchor provisioned in advance by the browser. Its presence is independent of which sites the user has visited, so the broker can return a verified target immediately. The second is cached data left after another site installs an LLM. Its presence may reflect browsing history, so the broker waits until the public deadline $D(p)$ before returning the result. Both classes use the same lookup and reconstruction process; they differ in planning objective and return time. Figure~\ref{fig:architecture} shows this common process and the return control used for strict release.

Steps 1--8 show the complete process. The page submits an installation request for the target LLM (1). Before querying shared cache state, the broker decides whether to admit the request. Under strict release, this step also sets $D(p)$ (2), after which the broker prepares the return control (3). Once the request is admitted, the broker privately finds eligible cached data (4), selects an action for each chunk (5), fetches the required objects (6), reconstructs and verifies the target (7), and returns the LLM handle according to the release policy (8). Public-anchor mode returns the handle as soon as the target is verified, whereas strict release waits until $D(p)$. The page cannot select the base package or observe the cache lookup, reconstruction plan, transfer volume, or execution progress.

\subsection{Target Verification and Package Alignment}

The publisher-signed manifest defines the target file together with its semantic regions and chunks, and the broker delivers only a result that exactly matches the signed content. The GGUF adapter validates the file structure, aligns packages from different publishers by tensor name and internal position, and divides large tensors into fixed 1-MiB chunks. Byte-identical chunks can be reused directly, aligned but different data uses residual reconstruction, and portions that cannot use equal-length reconstruction are obtained through compressed fetch. For each target chunk, the public catalog lists the permitted operations and binds each residual to its base package, target, transform, and data length.

\subsection{Per-Chunk Reconstruction Selection}

After verifying the target and aligning the model packages, \system reads the precomputed reconstruction operations that the public catalog lists for each target chunk and selects the lowest-cost one according to transfer volume, decoding time, and memory demand.

\textbf{Exact reuse and compressed fetch.} If a target chunk is identical to local data, \system uses \emph{exact reuse} and reads the bytes locally without network transfer. If local data cannot be reused efficiently, the system uses \emph{compressed fetch} and downloads the zstd-compressed target bytes from the publisher. Compressed fetch ensures that the system can always obtain the target when the other reconstruction actions do not reduce cost.

\textbf{XOR and BMS residuals.} For chunks with equal length but different contents, \system can transfer only the residual between the base package and target. Let $b_i$ be a base-package byte and $t_i$ the corresponding target byte. The publisher generates two basic residuals offline:
\[
  x_i=b_i\oplus t_i,
  \qquad
  s_i=(t_i-b_i)\bmod 256.
\]
Here $x_i$ is the \emph{XOR residual}. Applying XOR again during decoding recovers the target byte. The value $s_i$ is the \emph{Bytewise Modular Subtraction (BMS) residual} used by \system for GGUF reconstruction. The decoder recovers the target byte as
\[
  t_i=(b_i+s_i)\bmod 256.
\]
Bytewise subtraction has previously been used in binary differencing, the Portable Network Graphics (PNG) Sub filter, and integer-mapped LLM deltas~\cite{bsdiff,png-sub,fm-delta}. The key observation behind \system is that independent publishers often produce only small numeric changes when they requantize the same LLM with different calibration data. BMS consequently produces a more compressible residual for many aligned chunks. Other chunks follow different change patterns and can produce a smaller XOR residual. The publisher precomputes and compresses both residuals, and \system selects the one with the smaller complete transfer size for each chunk.

\textbf{Low-bit lane differences.} GGUF packs several 2- or 4-bit quantized values into one byte, so comparing the whole byte can hide relationships between the individual values. \emph{Low-bit lane differences} separate the packed 2-, 4-, or 8-bit values and compute a circular signed difference for each value. The result can also undergo bit-plane transposition before zstd compression~\cite{zstd-rfc,bitshuffle}. The decoder reverses these operations in the opposite order, so the target data remain unchanged.

\textbf{Field separation.} A quantized block commonly contains scales, subscales, and low-bit codes that represent LLM weights. These fields have different meanings and change patterns, so processing them as one byte stream can reduce compression effectiveness. For Q5\_0, Q4\_K, and Q6\_K, \system uses the validated GGML type to locate the fields and separates the scale, subscale, and packed-code streams. Each field can independently select exact reuse, compressed fetch, XOR, BMS, or a low-bit lane difference. A change in one field therefore does not prevent the other fields in the same block from being reused.

\textbf{Public-anchor planning.} Public anchors are provisioned by the browser and do not reflect which sites a user has visited, so the result can be returned as soon as reconstruction finishes. Among memory-feasible candidates, the planner selects the action with the shortest estimated delivery time for each chunk:
\[
  8(B_a+h)/r+d_a.
\]
Here $B_a$ is the action payload, $h$ is its framing overhead, $r$ is the network rate, and $d_a$ is the decode time. This cost model allows the planner to trade network transfer against local computation.

\textbf{Strict-release planning.} When \system reuses data created by another site's activity, the final handle cannot be returned until the public deadline $D(p)$. Finishing earlier therefore does not make the result available to the page sooner. The planner instead minimizes transferred bytes while ensuring that reconstruction finishes on time:
\[
  \min_{\pi}\sum_{a\in\pi} B_a
  \quad\text{s.t.}\quad
  \widehat T(\pi)\le D(p)-t_0,
  \quad M(\pi)\le M_{\max}.
\]
Here $t_0$ is the admission time, $\pi$ is the complete reconstruction plan, $\widehat T(\pi)$ is its estimated completion time, including metadata processing, network transfer, round trips, decoding, and reserved local processing, and $M(\pi)$ is its memory demand. The planner first chooses the lowest-traffic action for each chunk that satisfies the memory bound. If the complete plan cannot finish before $D(p)$, it replaces selected actions with faster alternatives and can fetch the complete target when necessary. The internal choice of plan never changes the return time observed by the page.

\subsection{Streaming Reconstruction and Delivery}

After selecting the reconstruction plan, the broker fetches data while reconstructing the target and verifies the complete file before delivering it to the page. Reconstruction uses a bounded stream and does not load the entire LLM file into memory. The broker reads only the required ranges of the base package, overlaps network transfer with reconstruction, and writes the result to a temporary file through synchronous OPFS access handles~\cite{fs-standard}. It verifies each chunk while writing, then rereads the complete file and verifies the final target against the publisher-authorized manifest. Only a file that matches exactly is committed and made available through the final handle. Public anchors are pinned and change only with browser updates, so their presence never depends on local storage pressure. The Rust/WebAssembly (Wasm)~\cite{wasm} kernel and download buffers use fixed memory, while per-site request limits, concurrent-install limits, and global quotas control cache occupancy.

We implemented a complete \system prototype comprising a privileged Chromium extension, a Rust/WebAssembly reconstruction kernel, and publishing tools. Cache lookup, plan selection, reconstruction, and verification all execute inside the browser broker. Appendix~\ref{app:implementation} gives the planning budgets, resource limits, commit procedure, and prototype implementation details.

\section{Performance Evaluation}\label{sec:eval}

The evaluation has two parts: offline evaluation and end-to-end browser experiments. The offline evaluation compares traffic savings across methods, analyzes why quantization residuals are compressible, tests the method on new publishers, new LLMs, and 7B LLMs, and measures reconstruction speed and resource overhead. The end-to-end browser experiments run the complete cross-site delivery workflow in Chromium and measure HTTP response-body bytes and page-ready time.

\subsection{Experimental Setup}

\textbf{Datasets.} The evaluation uses four datasets. The core dataset contains the 23 directed model-package pairs identified by the audit in Section~\ref{sec:corpus}, spanning 12 LLM-family/quantization groups. The independent validation set contains eight model packages covering two new LLMs and four publishers that were not involved in codec design. The 7B dataset consists of Qwen2.5-7B-Instruct Q4\_K\_M model packages from three publishers, forming six directed pairs with a 4.36-GiB target file in each direction. The controlled-quantization dataset fixes the 16-bit floating-point weights of SmolLM2 and the llama.cpp version while changing only the calibration data or tensor-type policy~\cite{gguf}. The first three datasets use publicly released GGUF model packages from Hugging Face~\cite{huggingface}. The end-to-end browser experiment selects one directed pair from each of the four LLM families in the core dataset and uses the same model-package sources.

\textbf{Comparators.} The comparators fall into three groups. The first represents ordinary LLM delivery: a level-3 zstd full download~\cite{zstd-rfc} and browser-partitioned whole-file caching~\cite{cache-partition}. The second consists of generic binary-reuse methods: FastCDC~\cite{fastcdc}, whole-file Xdelta, and tensor-aligned Xdelta~\cite{vcdiff-rfc,xdelta}. Tensor-aligned Xdelta receives the same tensor boundaries and complete equal-length base blocks as \system, testing whether \system's advantage comes only from tensor alignment. The third group uses LLM structure: a ZipLLM/BitX-style bit shuffle~\cite{zipllm}, exact tensor reuse, fixed tensor XOR, and fixed BMS. The last three are internal \system ablations that measure the contributions of exact reuse, XOR residuals, and BMS residuals, respectively. \system core selects among exact reuse, compressed fetch, XOR, and BMS for each chunk; full \system, evaluated in Section~\ref{sec:e2e}, additionally uses low-bit lane differences and field separation. Xdelta, XOR, and BMS may also use the same exact-reuse and level-3 zstd compressed-fetch operations, so their differences come from residual coding and operation selection.

\textbf{Environment and statistics.} Both the offline evaluation and the end-to-end browser experiments use only CPUs. Appendix~\ref{app:implementation} gives the complete experimental environments, comparator parameters, statistical methods, and reproduction description.

\subsection{Traffic Savings and Result Stability}\label{sec:traffic}

\system's core planner achieves the greatest traffic reduction in Table~\ref{tab:traffic}, which reports network traffic saved relative to a level-3 zstd full download. The table includes the envelopes transmitted by the reconstruction operations; manifest and catalog traffic is accounted for separately. Every pair in the full core dataset saves at least 63.3\%. Even when Xdelta receives the same tensor-alignment information, \system wins on all 23 pairs and leads by at least 11.6 percentage points on every pair. Fixed BMS outperforms fixed XOR, showing that subtraction residuals further reduce traffic. \system core selects a residual encoding for each chunk according to its contents and therefore obtains the best result in the table. Including manifests and catalogs reduces this result by only 0.3 percentage points.

\begin{table}[t]
  \caption{Median traffic reduction on the 23 calibration-divergent target pairs. Fixed-codec rows retain exact reuse and compressed fetch while fixing the delta transform.}
  \label{tab:traffic}
  \centering
  \small
  \begin{tabular}{@{}lr@{}}
    \toprule
    Method & Reduction (\%) \\
    \midrule
    Whole-file cache / full zstd & 0.0 \\
    FastCDC, 1-MiB average & 23.8 \\
    FastCDC, 256-KiB average & 23.9 \\
    Tensor-reset FastCDC & 24.0 \\
    Whole-file Xdelta & 30.1 \\
    Tensor-aligned Xdelta & 49.4 \\
    Tensor chunks, exact & 23.9 \\
    Tensor chunks + BitX & 68.6 \\
    Tensor XOR (fixed codec) & 75.4 \\
    BMS (fixed codec) & 76.4 \\
    \system core & \textbf{77.3} \\
    \bottomrule
  \end{tabular}
\end{table}

\system outperforms fixed XOR on all 23 pairs and all 12 LLM-family/quantization groups. When weighted by target-file size, \system saves 76.2\%, compared with 68.8\% for the strongest external comparator (BitX-style bit shuffle) and 51.6\% for tensor-aligned Xdelta. Appendix~\ref{app:publisher} reports the complete statistical results.

\textbf{Why BMS works.} BMS gains its advantage from the distribution of residual values. XOR compares the bits of two bytes, whereas BMS computes the target byte minus the base byte modulo 256. When quantized values change only slightly, BMS residuals usually remain close to zero and are therefore easier to compress. On aligned tensor chunks whose contents differ, compressed BMS residuals are 8.89\% smaller than XOR residuals, and BMS produces a smaller aggregate residual on all 23 pairs. For a nonzero residual byte $v$, define its circular magnitude as $\min(v,256-v)$. In BMS, 31.59\% of residual magnitudes are at most one, compared with 17.52\% in XOR. On the median pair, the 90th percentiles of the two residual-magnitude distributions are 74 and 99, respectively. These results show that BMS preserves numerical proximity between quantized values.

\begin{figure*}[t]
  \centering
  \includegraphics[width=\textwidth]{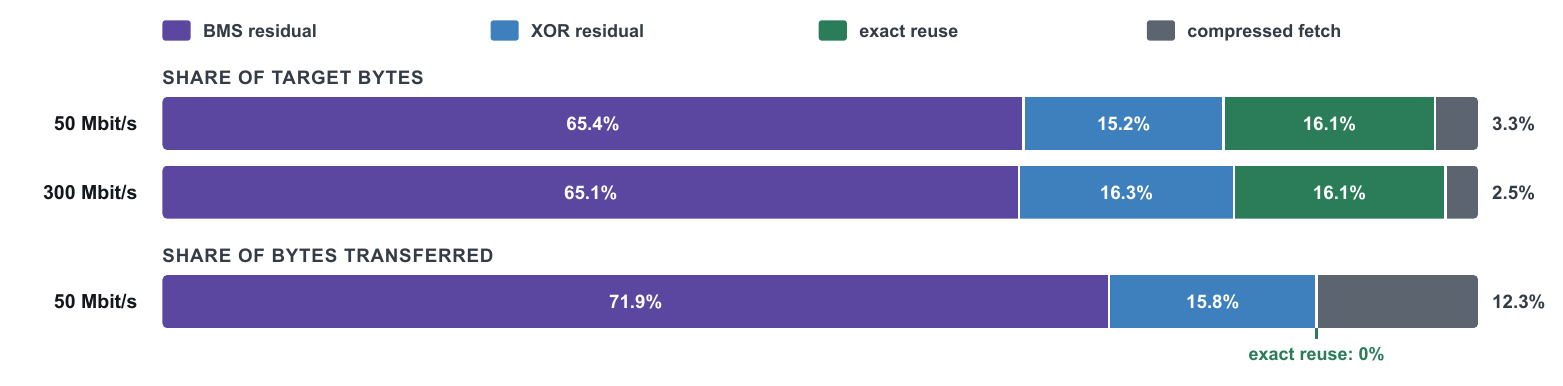}
  \caption{Operations selected by the latency-optimized policy across the 23 calibration-divergent pairs. The upper two bars show the share of target bytes handled by each operation at 50 and 300 Mbit/s. The lower bar shows each operation's share of transmitted bytes at 50 Mbit/s. Exact reuse transfers no network data.}
  \Description{Three stacked bars compare BMS residuals, XOR residuals, exact reuse, and compressed fetch at two bandwidths and show their shares of target and transmitted bytes.}
  \label{fig:planmix}
\end{figure*}

Figure~\ref{fig:planmix} shows the operations selected by the latency-optimized policy. BMS handles most tensor data that has changed but remains aligned, XOR handles chunks better suited to bitwise comparison, identical chunks are reused locally, and only portions that cannot be reconstructed efficiently are fetched from the publisher. As bandwidth increases, the planner adjusts the share assigned to each operation, showing that operation selection considers both network transfer and local decoding cost. The strict-release policy, which minimizes transferred bytes, achieves the same or slightly higher traffic savings.

\textbf{Publisher sensitivity.} To test whether the main result depends on any single publisher, the evaluation groups results by publisher combination and recomputes them after excluding each publisher in turn. \system retains the best traffic reduction throughout these analyses. Appendix~\ref{app:publisher} reports the complete results.

\subsection{Controlled Validation, Independent Validation, and 7B LLM Evaluation}

Building on the core-dataset results in Section~\ref{sec:traffic}, this section uses controlled quantization to validate the relationship between calibration differences and residual savings, then evaluates \system on new publishers, new LLMs, and larger LLMs.

\textbf{Controlled validation.} The controlled experiment fixes the LLM weights and quantization toolchain and changes only the calibration method or tensor-type policy, allowing the effects of these two differences on reconstruction to be observed separately. Appendix~\ref{app:scale-pairs} gives the complete model-package generation configuration and comparison directions.

When only calibration changes, tensor types, sizes, and positions remain fixed; only the quantized values change. Table~\ref{tab:controlled} shows that BMS produces smaller residuals than XOR in every calibration comparison, confirming that calibration differences create quantization changes suited to modular subtraction. When only the tensor-type policy changes, the storage format and length of some tensors also change. The planner fetches these tensors because equal-length residual coding cannot be applied, while exactly reusing tensors whose format is unchanged. The controlled experiment therefore validates the division of labor between the operations: BMS handles quantization changes that preserve layout, while fetch and exact reuse handle tensor-format changes.

\begin{table}[t]
  \caption{Controlled quantization with fixed weights and toolchain. Values are group medians over tensor bytes whose storage format and length remain unchanged, and BMS gain denotes residual reduction relative to XOR.}
  \label{tab:controlled}
  \centering\small
  \begin{tabular}{@{}llrrr@{}}
    \toprule
    Tensor policy & Calibration contrast & Directions & Bit flips & BMS gain \\
    \midrule
    Mixed & None--calibrated & 8 & 5.32\% & 7.33\% \\
    Mixed & Prose--code & 4 & 3.15\% & 7.08\% \\
    Pure & None--calibrated & 8 & 6.24\% & 8.60\% \\
    Pure & Prose--code & 4 & 3.91\% & 9.19\% \\
    \bottomrule
  \end{tabular}
\end{table}

\textbf{Independent validation.} The independent validation set contains eight model packages covering two new LLMs (Qwen3-0.6B and TinyLlama-1.1B-Chat-v1.0) and four publishers that were not involved in codec design. The four Qwen3-0.6B directions contain large amounts of identical tensor data, yielding 99.52--99.57\% exact reuse and 99.84--99.86\% traffic savings with \system. In the four TinyLlama-1.1B-Chat-v1.0 directions, nearly every chunk differs between publishers; \system saves 97.8\%, compared with 97.5\% for fixed XOR and 95.0\% for tensor-aligned Xdelta. These two groups validate \system's ability to reuse both identical tensor data and data that changes in nearly every chunk, showing that its benefits extend to publishers and LLMs outside the core dataset.

\textbf{7B LLM evaluation.} The 7B experiment uses Qwen2.5-7B-Instruct Q4\_K\_M model packages from three publishers. Each target file is 4.36 GiB, and the three model packages form six directed pairs. Table~\ref{tab:scale} shows that \system outperforms tensor-aligned Xdelta and fixed tensor XOR in all six directions, with both its median and minimum traffic savings reaching 76.6\%.

\begin{table}[t]
  \caption{Traffic savings on six cross-publisher directions for 7B LLMs. Values are relative to a level-3 zstd full download.}
  \label{tab:scale}
  \centering\small
  \begin{tabular}{@{}lrr@{}}
    \toprule
    Method & Median (\%) & Minimum (\%) \\
    \midrule
    Exact tensor reuse & 16.0 & 9.5 \\
    Tensor-aligned Xdelta & 43.9 & 43.9 \\
    Fixed tensor XOR & 73.1 & 73.1 \\
    \system core & \textbf{76.6} & \textbf{76.6} \\
    \bottomrule
  \end{tabular}
\end{table}

The six directions form two groups. In the four directions that include importance-matrix calibration differences, \system's BMS residuals are 13.2\% smaller than XOR residuals, and every direction saves 76.6\% of traffic. In the other two directions, neither package records an importance matrix, so more tensor data can be reused exactly and the remaining changes produce only small residuals, raising savings to 99.9\%. These results show that \system handles both calibration-induced quantization changes and local differences between publisher packages at 7B scale. Appendix~\ref{app:scale-pairs} reports the complete results for all six directions.

\subsection{Reconstruction Speed and Resource Overhead}

Reducing network traffic is useful in practice only if local reconstruction is fast enough and its memory and storage overheads remain controlled. This section evaluates \system's local reconstruction performance and its memory and storage overheads.

\textbf{Complete reconstruction performance.} Across all 42 pairs, \system achieves a median reconstruction throughput of 105.1 MiB/s. In repeated tests over six pairs covering all four LLM families, the median throughput is 196.2 MiB/s and maximum resident memory is 93.7 MiB. These results show that \system can reconstruct LLM files as a stream without loading the complete file into memory.

\textbf{Codec overhead.} With the same tensor alignment and equal-length base blocks, the median decoding time for the residuals selected by \system is 0.401 s, compared with 1.290 s for Xdelta, making \system about 3.2 times as fast.

\textbf{Catalog and storage scaling.} The 45 manifests and 42 catalogs occupy 28.55 MiB in total. Candidate residuals occupy 1.91 GiB, about 10\% of the 18.4-GiB corpus; routing reconstruction through a shared hub package cuts this storage by 37.8\% and adds only 2.9\% median traffic. Appendix~\ref{app:planning} reports the complete results.

\subsection{End-to-End Cross-Site Delivery}\label{sec:e2e}

\textbf{Purpose and setup.} The preceding experiments measure traffic savings and local reconstruction performance separately. This section tests whether these advantages shorten page waiting time during complete browser delivery. The experiment selects one cross-publisher Q4\_K\_M pair from each of the four LLM families in the core dataset, with target sizes ranging from 100.6 to 770.3 MiB. Each pair runs five times under full \system and each of the five comparator configurations described below, at both 50 Mbit/s and 100 Mbit/s with an 80-ms network round-trip time, producing 240 experiments. Each experiment starts with a public anchor already stored by the browser, namely the other publisher's package of the same model and quantization. Because one browser-provisioned anchor serves every site that requests a matching target, its download is excluded from the measured traffic. Each experiment measures the complete time from the page submitting a target to receiving a verified LLM, and counts all delivery data transferred during this process.

\textbf{Comparator configurations.} The experiment compares full \system with five configurations. \textbf{Ordinary partitioned full delivery} directly downloads the complete compressed target and measures the overall benefit of cross-publisher reuse. \textbf{Fixed tensor XOR} uses the same tensor alignment, exact reuse, and compressed fetch as \system, but uses XOR for every residual; it tests the value of selecting different reconstruction operations for individual chunks. Three additional configurations analyze the quantization-specific operations. \textbf{Basic actions} support exact reuse, compressed fetch, XOR, and BMS. \textbf{Basic actions plus low-bit lane differences} additionally process packed low-bit weights. \textbf{Field-separated XOR} separates scale values, subscale values, and weight codes, but uses XOR for every changed field. Full \system may use all operations and selects the plan with the shortest estimated delivery time for each chunk or field.

\begin{table}[t]
  \caption{Reductions of full \system relative to five browser comparator configurations. Values are medians [minimum, maximum] across eight model-package-pair/bandwidth combinations.}
  \label{tab:browser}
  \centering\scriptsize
  \setlength{\tabcolsep}{2pt}
  \renewcommand{\arraystretch}{1.05}
  \begin{tabular}{@{}>{\raggedright\arraybackslash}p{0.38\columnwidth}>{\centering\arraybackslash}p{0.27\columnwidth}>{\centering\arraybackslash}p{0.27\columnwidth}@{}}
    \toprule
    Comparator configuration & \shortstack{Page-ready time\\reduction (\%)} & \shortstack{HTTP response-body\\reduction (\%)} \\
    \midrule
    Ordinary partitioned full delivery & 64.2 [54.1, 77.5] & 78.9 [72.2, 86.5] \\
    Fixed tensor XOR & 12.6 [4.1, 19.1] & 23.0 [15.6, 33.1] \\
    Basic actions & 9.2 [3.7, 14.7] & 18.5 [12.5, 23.2] \\
    Basic actions plus low-bit lane differences & 7.9 [4.8, 12.9] & 14.2 [12.3, 22.1] \\
    Field-separated XOR & 9.6 [2.2, 15.5] & 15.0 [9.2, 28.8] \\
    \bottomrule
  \end{tabular}
\end{table}

\textbf{End-to-end results.} As Table~\ref{tab:browser} shows, relative to ordinary partitioned full delivery, \system reduces network traffic by 78.9\% and page-ready time by 64.2\%. Relative to fixed tensor XOR with the same tensor alignment, it further reduces network traffic by 23.0\% and page-ready time by 12.6\%. This shows that the traffic advantage in the offline evaluation translates into a real reduction in page waiting time. \system also outperforms the three internal ablations. The results show that low-bit lane differences and field separation each reduce delivery cost further, while selecting different residuals for different fields outperforms using XOR throughout. Table~\ref{tab:browser-cells} in Appendix~\ref{app:browser-cells} reports the complete results for each model-package-pair/bandwidth combination.

\section{Related Work}\label{sec:related}

\system relates to two bodies of work: Web data delivery and cross-site reuse, and reconstruction of a target from an existing file.

\textbf{Web delivery mechanisms.} Distributed Web caching studies object placement among servers~\cite{cache-placement-web}, and Hugging Face's Xet storage deduplicates model files at chunk level~\cite{hf-xet}, which, like FastCDC, recovers little from calibration-divergent packages. On the client side, HTTP delta encoding transfers differences from a cached version, but the client must disclose that version to the server~\cite{rfc3229}. Cross-Origin Storage (COS) and Cross-Origin Model Cache allow sites to reuse data but require byte-identical files~\cite{cos,hybrid-cache}. Compression Dictionary Transport (CDT) can reconstruct a byte-different HTTP response, but the site must preselect the dictionary and can use it only within that site's partition~\cite{chrome-cdt,cdt-rfc}. Chrome's built-in AI provisions one on-device model at the browser level~\cite{chrome-builtin-ai}, a precedent for browser-provisioned public anchors. \system lets the page specify only a publisher-authorized target, while the browser independently selects a byte-different cross-publisher base package and hides that choice from the page. Table~\ref{tab:rw-summary} summarizes this distinction.

\begin{table}[t]
  \caption{Browser delivery support for independently repackaged LLMs. \rwyes{} means supported; \rwout{} means unsupported.}
  \label{tab:rw-summary}
  \centering
  \scriptsize
  \setlength{\tabcolsep}{2.2pt}
  \renewcommand{\arraystretch}{1.02}
  \begin{tabular}{@{}>{\raggedright\arraybackslash}p{0.38\columnwidth}>{\centering\arraybackslash}p{0.16\columnwidth}>{\centering\arraybackslash}p{0.19\columnwidth}>{\centering\arraybackslash}p{0.20\columnwidth}@{}}
    \toprule
    Mechanism & \shortstack{Cross-site\\reuse} & \shortstack{Byte-different\\base} & \shortstack{Browser selects\\and hides base} \\
    \midrule
    HTTP delta encoding~\cite{rfc3229} & \rwout & \rwyes & \rwout \\
    Cross-Origin Storage (COS)~\cite{cos} & \rwyes & \rwout & \rwout \\
    Cross-Origin Model Cache~\cite{hybrid-cache} & \rwyes & \rwout & \rwout \\
    Compression Dictionary Transport (CDT)~\cite{cdt-rfc} & \rwout & \rwyes & \rwout \\
    \textbf{EvoCache} & \textbf{\rwyes} & \textbf{\rwyes} & \textbf{\rwyes} \\
    \bottomrule
  \end{tabular}
\end{table}

\textbf{Reconstruction methods.} LBFS and FastCDC use content-defined chunking to find identical data~\cite{lbfs,fastcdc}; VCDIFF, Xdelta, and bsdiff generate deltas for related binary files~\cite{vcdiff-rfc,xdelta,bsdiff}. LLM-aware methods such as ZipNN, FM-Delta, QStore, and ZipLLM use weight or tensor structure but generally require a fixed representation, a known training relationship, matched precision, or a uniformly managed model collection~\cite{zipnn,fm-delta,qstore,zipllm}. Independently published GGUF files may change metadata, calibration results, and tensor formats together and need not meet these assumptions. \system targets this setting by using tensor structure and quantization layout to select exact reuse, compressed fetch, or residual reconstruction for each chunk or field inside the browser. Tables~\ref{tab:rw-systems} and~\ref{tab:rw-algorithms} in Appendix~\ref{app:related-comparisons} give complete comparisons.

\section{Conclusion and Future Work}\label{sec:conclusion}

Cross-publisher GGUF repackaging of the same LLM prevents whole-file cache reuse, but corresponding tensors retain substantial compressible information. \system lets a page specify only an authorized target, while the browser internally selects a base package and a lossless reconstruction operation for each chunk. On the core dataset, \system saves 77.3\% of network traffic, exceeding Xdelta given the same tensor-alignment information (49.4\%); it still saves 76.6\% on 4.36-GiB 7B LLMs. Across 240 real end-to-end browser experiments, \system reduces response-body bytes by 78.9\% and page-ready time by 64.2\% relative to ordinary full delivery. Results reconstructed from public anchors can be returned immediately, whereas history-related cache reuse follows a public deadline, thereby protecting cross-site cache privacy. Deployment requires little publisher effort: catalogs are static objects generated offline, and candidate residuals add about 10\% storage that a hub package further reduces. Browsers can choose anchors from widely shared LLM families and bound anchor storage through browser provisioning policy. Although we evaluate GGUF, the same tensor alignment and per-chunk selection apply to other block-quantized formats. Future work includes popularity-driven anchor selection, cache-state-oblivious prefetching, private data retrieval, and multi-base selection.

\bibliographystyle{ACM-Reference-Format}
\bibliography{references}

\appendix
\section{Conditions Required by Strict Release}\label{app:proof}

Section~\ref{sec:release}'s conditional API equality requires: (1) admission, the release rule, and deadlines read only public request classes and caller-owned quota, never broker occupancy; (2) the shared store has an accounting domain independent of caller storage; (3) reconstruction requests and progress are absent from caller interception and Performance APIs; (4) both accepted paths complete by the deadline; and (5) neither path emits its handle or error before that time. Opaque handles are sampled independently of cache state and bound to the originating caller.

Under these conditions, couple the two executions' public admission, random handle generation, and release schedule. Their intermediate API sequences are empty, and their terminal values describe the same target at the same scheduled time, establishing equality in distribution of the API transcripts. Extending this equality to arbitrary page observations additionally requires cache-state-independent CPU, storage, and network service classes, which we leave to browser-level resource isolation.

\section{Implementation and Evaluation Details}\label{app:implementation}

\textbf{System implementation details.} The \system prototype comprises a Python reference implementation, a Rust/Wasm reconstruction kernel, a memory-bounded native Xdelta wrapper, and a privileged Chromium extension. In Chromium 151, the extension implements the browser broker using an offscreen document, a hidden worker, OPFS storage, and persistent model-package and chunk indexes. Pages cannot inspect the internal indexes or specify the base package or reconstruction plan. The publisher generates the manifest, content-addressed chunks, and catalog offline and signs the manifest. Median packaging and catalog-generation times are 3.45 and 6.77 s, respectively. The generated static objects can be uploaded alongside ordinary LLM files, without online LLM execution or a specialized delivery server.

Across 336 catalog and network combinations, the strict-release plan that prioritizes reducing transferred bytes satisfies the public time budget in every case. At 300 Mbit/s, its traffic reduction differs from that of the latency-optimized policy by 0.040 percentage points, with a maximum difference of 0.171 percentage points. The Rust/Wasm kernel uses a fixed 16-MiB memory region to hold one base-package chunk, the decoded residual, and the target chunk. Remote bundles do not exceed 8 MiB in either compressed or decoded form, and at most two decoded bundles are retained simultaneously. WebCrypto checks publisher authorization, the reconstructed result is written to a temporary OPFS file, and it is committed after verification. Per-site and global quotas jointly limit cache occupancy, and shared broker storage is accounted for separately.

\textbf{Experimental and reproduction details.} The offline experiments, except the 7B evaluation, run on a Windows machine with an Intel Core i9-11900K processor and 32 GiB of memory. The 7B evaluation and the end-to-end browser experiments run on a Linux 6.8 virtual machine in Melbourne Research Cloud with eight AMD EPYC 7763 virtual CPUs and 32 GiB of memory. All codecs use only CPUs. Model-package versions, target files, tensor chunking, and comparator configurations were fixed before the full experiment and kept unchanged across the core dataset and independent validation set. Each directed pair constitutes one observation. The 95\% confidence intervals use 10,000 cluster-bootstrap resamples over 12 LLM-family/quantization groups~\cite{bootstrap}, and paired Wilcoxon tests compare \system with fixed XOR~\cite{wilcoxon}. FastCDC was tested with average chunk sizes of 256 KiB, 1 MiB, and 4 MiB and region-reset settings of 64 KiB, 256 KiB, and 1 MiB. Xdelta 3.2.0 uses its maximum compression level (\texttt{-9}) with DJW secondary compression; whole-file Xdelta uses a 16-MiB source window, the largest setting that keeps the smallest packages within a memory budget of twice the model size, and tensor-aligned Xdelta encodes each aligned block with a 4-MiB window. The end-to-end browser experiments use two independent HTTPS sites and one publisher server, with a fresh browser profile for every trial. The server provides all signed candidate objects but does not participate in selecting the base package or reconstruction plan. Across the complete process tree, peak resident memory is 1.42 GiB at the median and 1.49 GiB at maximum. We will release the \system artifact after the review process. Every reconstructed result exactly matches the requested publisher model package.

\section{Statistical Results and Publisher-Pair Sensitivity Analysis}\label{app:publisher}

The 95\% cluster-bootstrap confidence intervals are 66.5--78.8\% for fixed XOR, 67.8--79.0\% for fixed BMS, and 67.8--79.8\% for \system. \system outperforms fixed XOR on all 23 pairs, with a paired-test result of $p=1.19\times10^{-7}$. Its median is also higher in all 12 LLM-family/quantization groups containing these pairs, with $p=2.44\times10^{-4}$.

The publisher bartowski appears in 20 of the 23 target pairs; Table~\ref{tab:publisher-pairs} reports publisher-pair results, followed by leave-one-publisher-out results.

\begin{table}[H]
  \caption{Target-workload publisher-pair medians (traffic reduction, \%), with one representative of each comparator group in Table~\ref{tab:traffic}; $n=1$ is descriptive.}
  \label{tab:publisher-pairs}
  \centering
  \resizebox{\columnwidth}{!}{%
  \begin{tabular}{@{}lrrrrr@{}}
    \toprule
    Publisher pair & $n$ & \system & Fixed XOR & Aligned Xdelta & Tensor-reset FastCDC \\
    \midrule
    bartowski--lmstudio & 1 & 82.4 & 81.1 & 58.7 & 36.4 \\
    bartowski--mradermacher & 11 & 77.6 & 75.4 & 49.4 & 27.5 \\
    bartowski--unsloth & 8 & 73.4 & 72.3 & 51.5 & 20.3 \\
    mradermacher--unsloth & 3 & 73.1 & 70.4 & 48.3 & 3.0 \\
    \bottomrule
  \end{tabular}%
  }
\end{table}

Omitting bartowski/lmstudio/mradermacher/unsloth gives \system medians of 73.1\% ($n=3$), 77.2\% ($n=22$), 77.3\% ($n=9$), and 78.1\% ($n=12$). In the bartowski-disjoint subset, \system's 73.1\% savings alongside 3.0\% for tensor-reset FastCDC demonstrate positional-delta reuse even with little exact chunk overlap.

\section{Controlled-Experiment Configuration and Per-Direction 7B Results}\label{app:scale-pairs}

\textbf{Controlled-experiment configuration.} Real model packages may differ simultaneously in calibration data, tensor types, and file layout, making it difficult to determine the effect of each difference on reconstruction directly. The controlled experiment fixes the 16-bit floating-point weights of SmolLM2-135M and SmolLM2-360M and the llama.cpp b10816 toolchain, changing only the calibration method or tensor-type policy. For each LLM, it uses no importance matrix, a text-calibrated matrix, and a code-calibrated matrix under both pure-type and mixed-type policies, producing 12 model packages. Calibration processes 16 chunks of 128 tokens on the CPU. These model packages form 24 directions that change only calibration and 12 directions that change only the tensor-type policy. Table~\ref{tab:scale-pairs} gives the per-direction 7B results.

\section{Catalog Scaling}\label{app:planning}

For each LLM-family/quantization group, a public hub reduces pairwise catalogs from quadratic to linear edge count; a client reconstructs the hub and then the target. Across the 16 groups covering all 42 pairs, minimizing incident candidate bytes selects bartowski, mradermacher, and unsloth in seven, eight, and one groups. Edges fall from 42 to 29, candidate storage falls 37.8\%, and thirteen two-hop routes add 2.9\% median traffic.

Summed measured local reconstruction times are 3.37 s median and 17.10 s maximum for two-hop routes, or 1.24$\times$ direct CPU time at the median. Two-hop reconstruction temporarily needs 0.54 GiB median and 2.46 GiB maximum additional disk space for the intermediate hub and the target; including the existing base, the maximum is 3.69 GiB. The replay uses observed directed edges as reverse proxies, preserving XOR payload sizes and estimating directed BMS residuals, metadata, and timing.

\clearpage
\onecolumn
\begin{table}[H]
  \caption{Direction-level traffic reduction (\%) in the 7B evaluation. B/M/L denote bartowski, mradermacher, and lmstudio-community.}
  \label{tab:scale-pairs}
  \centering\scriptsize
  \begin{tabular}{@{}lrrrr@{}}
    \toprule
    Direction & Exact & Aligned Xdelta & Fixed XOR & \system \\
    \midrule
    B$\rightarrow$M & 9.538 & 43.924 & 73.068 & \textbf{76.614} \\
    M$\rightarrow$B & 9.538 & 43.910 & 73.081 & \textbf{76.626} \\
    B$\rightarrow$L & 15.982 & 43.936 & 73.077 & \textbf{76.623} \\
    L$\rightarrow$B & 15.979 & 43.923 & 73.091 & \textbf{76.635} \\
    M$\rightarrow$L & 26.979 & 99.861 & 99.896 & \textbf{99.899} \\
    L$\rightarrow$M & 26.980 & 99.861 & 99.896 & \textbf{99.899} \\
    \bottomrule
  \end{tabular}
\end{table}

\section{Browser-Combination Results}\label{app:browser-cells}

\begin{table}[H]
  \caption{Medians for each setting after five repetitions. Page time runs from submission of the target request until the final handle is returned; savings are relative to fixed tensor XOR, and response-body reduction includes signed control metadata.}
  \label{tab:browser-cells}
  \centering\scriptsize
  \begin{tabular}{@{}lrrrrrr@{}}
    \toprule
    Target & Mbit/s & Ordinary (s) & \system (s) & XOR (s) & Page save & Body save \\
    \midrule
    Llama-3.2-1B & 50 & 131.4234 & 33.3618 & 40.3131 & 17.243\% & 33.125\% \\
    Llama-3.2-1B & 100 & 69.0424 & 24.3758 & 28.3544 & 14.032\% & 33.123\% \\
    Qwen2.5-0.5B & 50 & 64.7999 & 14.5626 & 18.0101 & 19.142\% & 28.797\% \\
    Qwen2.5-0.5B & 100 & 34.1861 & 10.8640 & 12.2176 & 11.079\% & 28.745\% \\
    SmolLM2-135M & 50 & 17.8674 & 6.4882 & 7.5712 & 14.304\% & 15.585\% \\
    SmolLM2-135M & 100 & 9.7031 & 4.4313 & 4.6221 & 4.128\% & 15.585\% \\
    SmolLM2-360M & 50 & 44.9995 & 16.9995 & 19.0816 & 10.912\% & 17.237\% \\
    SmolLM2-360M & 100 & 24.1063 & 11.0713 & 12.0181 & 7.878\% & 17.237\% \\
    \bottomrule
  \end{tabular}
\end{table}

\section{Complete Related-Work Comparisons}\label{app:related-comparisons}

\begin{table}[H]
  \caption{Support for independently repackaged LLMs across browser delivery mechanisms.}
  \label{tab:rw-systems}
  \centering
  \scriptsize
  \setlength{\tabcolsep}{1.8pt}
  \renewcommand{\arraystretch}{1.00}
  \begin{tabular}{@{}>{\raggedright\arraybackslash}p{0.17\textwidth}>{\raggedright\arraybackslash}p{0.145\textwidth}>{\raggedright\arraybackslash}p{0.185\textwidth}>{\centering\arraybackslash}p{0.08\textwidth}>{\centering\arraybackslash}p{0.09\textwidth}>{\centering\arraybackslash}p{0.09\textwidth}>{\raggedright\arraybackslash}p{0.15\textwidth}@{}}
    \toprule
    Mechanism & Page request & Reusable data & \shortstack{Cross-site\\reuse} & \shortstack{Byte-different\\base} & \shortstack{Base hidden\\from page} & Release behavior \\
    \midrule
    HTTP delta encoding~\cite{rfc3229} & HTTP response & Client-declared cached version & No & Yes & No & Immediate \\
    Cross-Origin Storage (COS)~\cite{cos} & Public object identifier & Byte-identical public object & Yes & No & No & Immediate \\
    Cross-Origin Model Cache~\cite{hybrid-cache} & LLM identifier & Byte-identical complete LLM file & Yes & No & No & Immediate \\
    Compression Dictionary Transport (CDT)~\cite{cdt-rfc} & HTTP response & Prearranged, site-partitioned dictionary & No & Yes & No & Immediate \\
    \textbf{EvoCache} & \textbf{Publisher-authorized target} & \textbf{Public anchor or eligible cross-site package} & \textbf{Yes} & \textbf{Yes} & \textbf{Yes} & \textbf{Public anchor: immediate; history-derived cache: public deadline} \\
    \bottomrule
  \end{tabular}
\end{table}

\begin{table}[H]
  \caption{File relationships and decision units supported by reconstruction methods.}
  \label{tab:rw-algorithms}
  \centering
  \scriptsize
  \setlength{\tabcolsep}{2.0pt}
  \renewcommand{\arraystretch}{0.94}
  \begin{tabular}{@{}>{\raggedright\arraybackslash}p{0.105\textwidth}>{\raggedright\arraybackslash}p{0.15\textwidth}>{\raggedright\arraybackslash}p{0.19\textwidth}>{\raggedright\arraybackslash}p{0.245\textwidth}>{\raggedright\arraybackslash}p{0.26\textwidth}@{}}
    \toprule
    Method & Decision unit & Information used & Applicable source--target relation & Reconstruction method \\
    \midrule
    LBFS~\cite{lbfs} & Content-defined chunk & Byte content & Files retaining identical regions shifted by insertions & Exact-chunk lookup \\
    FastCDC~\cite{fastcdc} & Content-defined chunk & Byte content & Files containing identical chunks & Fast boundary selection \\
    Xdelta~\cite{xdelta} & Byte sequence & Generic byte information & Related binary files & VCDIFF copy/add operations~\cite{vcdiff-rfc} \\
    bsdiff~\cite{bsdiff} & Byte range & Generic byte information & Versions of the same software & Suffix matching and byte differences \\
    ZipNN~\cite{zipnn} & Weight block & Tensor storage representation & LLMs using one representation & Fixed lossless codec \\
    FM-Delta~\cite{fm-delta} & Parameter tensor & Numeric parameter type & Pretrained and fine-tuned LLMs & Integer-mapped parameter differences \\
    QStore~\cite{qstore} & Quantization group & Precision correspondence & Matched high- and low-precision LLMs & Quantization-conditioned residual \\
    ZipLLM~\cite{zipllm} & Tensor & Tensor identity and layout & Related LLMs in a repository & Tensor deduplication and BitX coding \\
    \textbf{EvoCache} & \textbf{Authenticated GGUF chunk or field} & \textbf{Tensor structure and quantization layout} & \textbf{The same LLM independently repackaged by different publishers} & \textbf{Lowest-cost exact-reuse, compressed-fetch, or residual action} \\
    \bottomrule
  \end{tabular}
\end{table}

\end{document}